\documentclass[twocolumn,floatfix,aps,superscriptaddress,nofootinbib]{revtex4-2}
\usepackage{graphicx}
\usepackage{amsmath}
\usepackage{amssymb}
\usepackage{bm}
\usepackage{hyperref}

\begin{document}

\title{Method to preserve the chiral-symmetry protection of the zeroth Landau level\\ 
on a two-dimensional lattice}
\author{A. Don\'{i}s Vela}
\affiliation{Instituut-Lorentz, Universiteit Leiden, P.O. Box 9506, 2300 RA Leiden, The Netherlands}
\author{G. Lemut}
\affiliation{Instituut-Lorentz, Universiteit Leiden, P.O. Box 9506, 2300 RA Leiden, The Netherlands}
\author{J. Tworzyd{\l}o}
\affiliation{Faculty of Physics, University of Warsaw, ul.\ Pasteura 5, 02--093 Warszawa, Poland}
\author{C. W. J. Beenakker}
\affiliation{Instituut-Lorentz, Universiteit Leiden, P.O. Box 9506, 2300 RA Leiden, The Netherlands}
\date{September 2022}
\begin{abstract}
The spectrum of massless Dirac fermions on the surface of a topological insulator in a perpendicular magnetic field $B$ contains a $B$-independent ``zeroth Landau level'', protected by chiral symmetry. If the Dirac equation is discretized on a lattice by the method of ``Wilson fermions'', the chiral symmetry is broken and the zeroth Landau level is broadened when $B$ has spatial fluctuations. We show how this lattice artefact can be avoided starting from an alternative  nonlocal discretization scheme introduced by Stacey. A key step is to spatially separate the states of opposite chirality in the zeroth Landau level, by adjoining $+B$ and $-B$ regions.
\end{abstract}
\maketitle

\section{Introduction}
\label{sec_intro}

\subsection{Objective}
\label{sec_objective}

We address, in a different context, a problem originating from lattice gauge theory: How to place fermions on a lattice in a way that respects both gauge invariance and chiral symmetry \cite{Nie81,Wil74,Kog75,Sus77,Sta82,Sta83,Gri93,Tong}. Our context is topological insulators \cite{Ber13}, three-dimensional (3D) materials having an insulating bulk and a conducting surface, with massless Dirac fermions as the low-energy excitations. The Landau level spectrum of massless Dirac fermions is anomalous, the zeroth Landau level is  a flat band pinned to zero energy irrespective of the magnetic field strength \cite{Rab28,McC56}.

Our objective is to model the surface states on a two-dimensional (2D) lattice, without breaking the chiral symmetry that protects the zeroth Landau level from broadening by disorder. Let us introduce the problem in some detail.

\subsection{Zeroth Landau level}
\label{sec_intro_LL}

In a magnetic field $B$, perpendicular to the surface of the topological insulator, Landau levels form at energies $E_n=\pm\hbar\omega\sqrt n$,\;\;$n\in\mathbb{N}$, with $\omega\propto\sqrt{B}$. The zeroth Landau level $E_0=0$ is magnetic-field independent \cite{Che10,Han10,Jia12,Cho20}. If the perpendicular field strength has spatial fluctuations, for example, because of ripples on the surface, all Landau levels are broadened \textit{except} the zeroth Landau level \cite{Gie07}. 

The $E=0$ flat band is protected by a chiral symmetry, a unitary and Hermitian operator ${\cal C}$ that anti-commutes with the Hamiltonian \cite{Chi16}. Indeed, the massless 2D Dirac Hamiltonian
\begin{equation}
H_{\rm D}=v\hbar k_x\sigma_x+v\hbar k_y\sigma_y \label{HDdef}
\end{equation}
anticommutes with the Pauli matrix $\sigma_z$, and this symmetry is preserved if one introduces a space-dependent vector potential by $\hbar\bm{k}\mapsto \hbar\bm{k}-e\bm{A}(\bm{r})$. 

Topological considerations \cite{Aha79,Alv83,Kat12} then enforce the existence of an ${\cal N}$-fold degenerate eigenstate at $E=0$, with ${\cal N}$ the number of flux quanta through the surface. The flat band has a definite chirality, meaning that it is an eigenstate of ${\cal C}=\sigma_z$ with eigenvalue $\pm 1$ determined by the sign of the magnetic field.

\begin{figure}[tb]
\centerline{\includegraphics[width=0.8\linewidth]{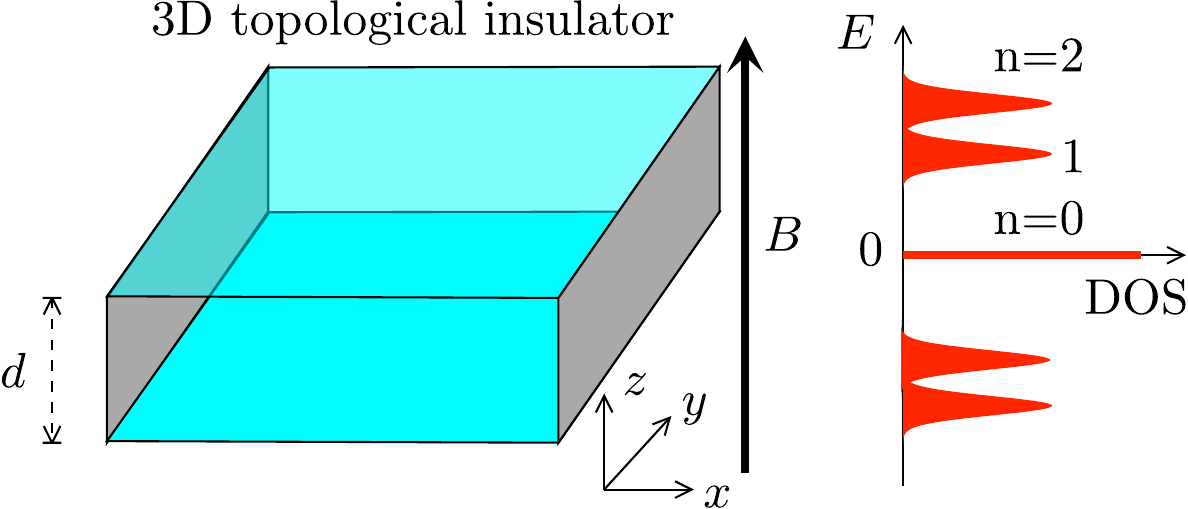}}
\caption{Slab of a topological insulator in a perpendicular magnetic field $B$. Landau levels form on the top and bottom surface at energy $|E|\propto\sqrt n$, $n=0,1,2,\ldots$, symmetrically arranged around $E=0$. The density of states (DOS) of the zeroth Landau level is not broadened by a spatially fluctuating $B$, provided that the slab thickness $d$ is sufficiently large that the two surfaces are decoupled.
}
\label{fig_layout}
\end{figure}

If we consider a topological insulator in the form of a slab (see Fig.\ \ref{fig_layout}), the top and bottom surfaces each support a zeroth Landau level, of opposite chirality. The two flat bands will mix and split if the slab is so thin that the wave functions of opposite surfaces overlap, but in thick slabs this breakdown of the topological protection is exponentially small in the ratio of slab thickness and penetration depth.

\subsection{2D lattice formulation}
\label{sec_intro_2D}

A numerical simulation of the 3D system is costly, it would be more efficient to retain only the surface degrees of freedom. If we discretize the 2D surface on a square lattice (lattice constant $a$), the Hamiltonian must be periodic in the momentum components with period $2\pi/a$. The $\sin ak$ dispersion has the proper periodicity, but it suffers from fermion doubling \cite{Tong}: a spurious massless degree of freedom appears at $k=\pi/a$. 

We contrast two lattice formulations that avoid fermion doubling: an approach due to Wilson \cite{Wil74} with a sine+cosine dispersion, and an approach due to Stacey \cite{Sta82} with a tangent dispersion.

In Wilson's approach \cite{Wil74} the discretized Dirac Hamiltonian is
\begin{align}
H_{\rm Wilson}={}&(\hbar v/a)\sum_{\alpha=x,y}\sigma_\alpha\sin ak_\alpha\nonumber\\
&+\Delta\sigma_z\sum_{\alpha=x,y}(1-\cos ak_\alpha).\label{HWdef}
\end{align}
The cosine term $\propto \Delta\sigma_z$ avoids fermion doubling, the only low-energy excitations are near $\bm{k}=0$, but it breaks chiral symmetry: $H_{\rm Wilson}$ no longer anticommutes with ${\cal C}=\sigma_z$.

The alternative approach due to Stacey \cite{Sta82} has a tangent dispersion,
\begin{equation}
H_{\rm Stacey}=(2\hbar v/a)\sum_{\alpha=x,y}\sigma_\alpha\tan(ak_\alpha/2).\label{HStaceydef}
\end{equation}
Fermion doubling is avoided without breaking chiral symmetry, at the expense of a nonlocal Hamiltonian: While sines and cosines of momentum only couple nearest neighboring sites, the tangent of momentum represents a long-range coupling.

The merit of Stacey's approach is that the nonlocal Schr\"{o}dinger equation $H_{\rm Stacey}\Psi=E\Psi$ can be cast in the form of a \textit{generalized} eigenvalue problem, 
\begin{equation}
{\cal H}\Psi=E{\cal P}\Psi,\label{genevproblem}
\end{equation}
with \textit{local} operators ${\cal H},{\cal P}$ given by \cite{Pac21}
\begin{subequations}
\label{HPdef}
\begin{align}
{\cal H}={}&\frac{\hbar v}{2a}\sigma_x(1+\cos ak_y)\sin ak_x\nonumber\\
&+\frac{\hbar v}{2a}\sigma_y(1+\cos ak_x)\sin ak_y,\\
{\cal P}={}&\tfrac{1}{4}(1+\cos ak_x)(1+\cos ak_y).
\end{align}
\end{subequations}
Because ${\cal H}$ and ${\cal P}$ are sparse Hermitian operators, and ${\cal P}$ is positive definite,\footnote{The operator ${\cal P}$ is in general only positive semidefinite. It becomes positive definite if we choose an odd number of lattice points with periodic boundary conditions in the $x$-- and $y$--directions.} the generalized eigenvalue problem \eqref{genevproblem} can be solved efficiently.

\subsection{Outline}
\label{sec_outline}

We wish to show that the topological protection of the zeroth Landau in a 3D topological insulator can be obtained in a purely 2D formulation. To preserve chiral symmetry we work with the tangent dispersion, in the local representation \eqref{HPdef}.

The first step is to introduce the vector potential in a gauge invariant way --- without breaking the locality of the generalized eigenvalue problem. We do this in the next Section \ref{sec_gaugeinvariant}. In Sec.\ \ref{sec_bandstructure} we calculate the Landau level spectrum. The zeroth Landau level contains states of both chiralities, we show that these can be spatially separated by adjoining $+B$ and $-B$ regions. The robustness of the flat band is assessed in Sec.\ \ref{sec_robustness}. We conclude in Sec.\ \ref{sec_conclude}.

\section{Gauge invariant lattice fermions with a tangent dispersion}
\label{sec_gaugeinvariant}

In Ref.\ \onlinecite{Pac21} it was shown how the magnetic field can be incorporated in the generalized eigenvalue problem \eqref{HPdef} in a way that is gauge invariant to first order in the flux through a unit cell. Here we will go beyond that calculation, and preserve gauge invariance to all orders. 

For ease of notation we set $\hbar$ and the lattice constant $a$ both equal to unity in most equations that follow. The electron charge is taken as $+e$, so that the vector potential enters in the Hamiltonian as $\bm{k}\mapsto \bm{k}-e\bm{A}$.

We recall the definition of the translation operator,
\begin{equation}
T_\alpha\equiv e^{i\hat{k}_\alpha}=\sum_{\bm{n}}|\bm{n}\rangle\langle\bm{n}+\bm{e}_\alpha|.\label{Talphadef}
\end{equation}
The sum over $\bm{n}$ is a sum over lattice sites on the 2D square lattice, and $\bm{e}_\alpha\in\{\bm{e}_x,\bm{e}_y\}$ is a unit vector in the $\alpha$-direction. The Peierls substitution ensures gauge invariance by the replacement
\begin{equation}
\begin{split}
&T_\alpha\mapsto {\cal T}_\alpha=\sum_{\bm{n}}e^{i\phi_\alpha(\bm{n})}|\bm{n}\rangle\langle\bm{n}+\bm{e}_\alpha|,\\
&\phi_\alpha(\bm{n})=e\int_{\bm{n}+\bm{e}_\alpha}^{\bm{n}}A_\alpha(\bm{r})\, dx_\alpha.
\end{split}\label{Peierls}
\end{equation}
Note that the $A$-dependent translation operators no longer commute,
\begin{equation}
{\cal T}_y{\cal T}_x=e^{2\pi i\varphi/\varphi_0}{\cal T}_x{\cal T}_y,
\end{equation}
where $\varphi$ is the flux through a unit cell in units of the flux quantum $\varphi_0=h/e$.

One could now apply the Peierls substitution directly to the Hamiltonian $H_{\rm Stacey}$ from Eq.\ \eqref{HStaceydef}, but then one runs into the obstacle noted in Ref.\ \onlinecite{Pac21}: The transformation to a local generalized eigenvalue problem only succeeds to first order in $A$, higher order terms become nonlocal. Here we therefore follow a different route.

We rewrite the operators ${\cal H}$ and ${\cal P}$ from Eq.\ \eqref{HPdef} in terms of the translation operators \eqref{Talphadef} and apply the Peierls substitution \eqref{Peierls} at that level. Noting that $1+\cos k_\alpha=\tfrac{1}{2}(1+T_\alpha)(1+T_\alpha^\dagger)$, $\sin k_\alpha=\tfrac{1}{2i}(T_\alpha-T^\dagger_\alpha)$, we define
\begin{subequations}
\label{HPdefcalT}
\begin{align}
{\cal H}={}&\frac{\hbar v}{8ia}\sigma_x(1+{\cal T}_y)({\cal T}_x-{\cal T}_x^\dagger)(1+{\cal T}_y^\dagger)\nonumber\\
&+\frac{\hbar v}{8ia}\sigma_y(1+{\cal T}_x)({\cal T}_y-{\cal T}_y^\dagger)(1+{\cal T}_x^\dagger),\\
{\cal P}={}&\Phi\Phi^\dagger,\\
\Phi={}&\tfrac{1}{8}(1+{\cal T}_x)(1+{\cal T}_y)+\tfrac{1}{8}(1+{\cal T}_y)(1+{\cal T}_x).
\end{align}
\end{subequations}

Since ${\cal T}_x$ and ${\cal T}_y$ do not commute the order matters: In Eq.\ \eqref{HPdefcalT} we have ordered these translation operators such that ${\cal H}$ and ${\cal P}$ remain Hermitian, and moreover ${\cal P}$ remains positive definite. Both these properties are essential for an efficient solution of the generalized eigenvalue problem
\begin{equation}
{\cal H}\Psi=E\Phi\Phi^\dagger\Psi.\label{calHPhiPhidag}
\end{equation}

For completeness we note that a scalar potential $V$ and a magnetization $\bm{M}$ can be included by adding to ${\cal H}$ the terms
\begin{equation}
{\cal H}\mapsto{\cal H}+\Phi V\Phi^\dagger+\Phi (\bm{M}\cdot\bm{\sigma})\Phi^\dagger.
\end{equation}
The potential $V$ and perpendicular magnetization $M_z$ break chiral symmetry, while the parallel magnetizations $M_x$ and $M_y$ preserve it.

\section{Chirality-resolved zeroth Landau level}
\label{sec_bandstructure}

\subsection{Lattice obstruction to chirality polarization}
\label{sec_obstruction}

The Landau levels of the Dirac Hamiltonian \eqref{HDdef} are dispersionless flat bands at energies $\pm E_n$ given by \cite{Rab28,McC56}
\begin{equation}
E_n=v\sqrt{2n\hbar e|B|},\;\;n=0,1,2,\ldots.\label{Endef}
\end{equation}
Each Landau level has the same degeneracy ${\cal N}$ = number of flux quanta through the system. Both chiralities $C=\pm 1$ (eigenvalues of $\sigma_z$) contribute equally to each nonzero Landau level: $\langle n|\sigma_z|n\rangle=0$ for $n\geq 1$. The zeroth Landau level, however, is polarized: $\langle 0|\sigma_z|0\rangle={\rm sign}\,B.$

The topological protection of the zeroth Landau level rests on this chirality polarization: The chirality index ${\cal I}$ of the zero-mode, equal to the number of states with $C=+1$ minus the number of states with $C=-1$, is equal to ${\cal I}=(\text{sign}\, B){\cal N}$. If chiral symmetry is maintained the index is a topological invariant \cite{Aha79,Alv83,Kat12}, preventing a broadening of the flat band.

All of this is for the continuum description. The fundamental obstacle faced by lattice fermions is that the chirality polarization of the zeroth Landau level is lost: A no-go theorem by Stacey \cite{Sta83} enforces that any gauge invariant lattice regularization of the Dirac Hamiltonian which preserves chiral symmetry must have the same number of zero-modes for either chirality. Hence, on the lattice ${\cal I}=0$ and the topological protection breaks down.

That gauge invariance on a lattice is incompatible with a nonzero chirality index might be understood by a topological argument: A uniform magnetic field can be concentrated into an array of $h/e$ flux tubes, each of which is fully contained within a unit cell. The chirality index cannot change by such a smooth deformation, but the resulting magnetic field distribution may be gauged away on the lattice, hence ${\cal I}$ must be equal to zero.

\subsection{Proposed work-around}

\begin{figure}[tb]
\centerline{\includegraphics[width=0.8\linewidth]{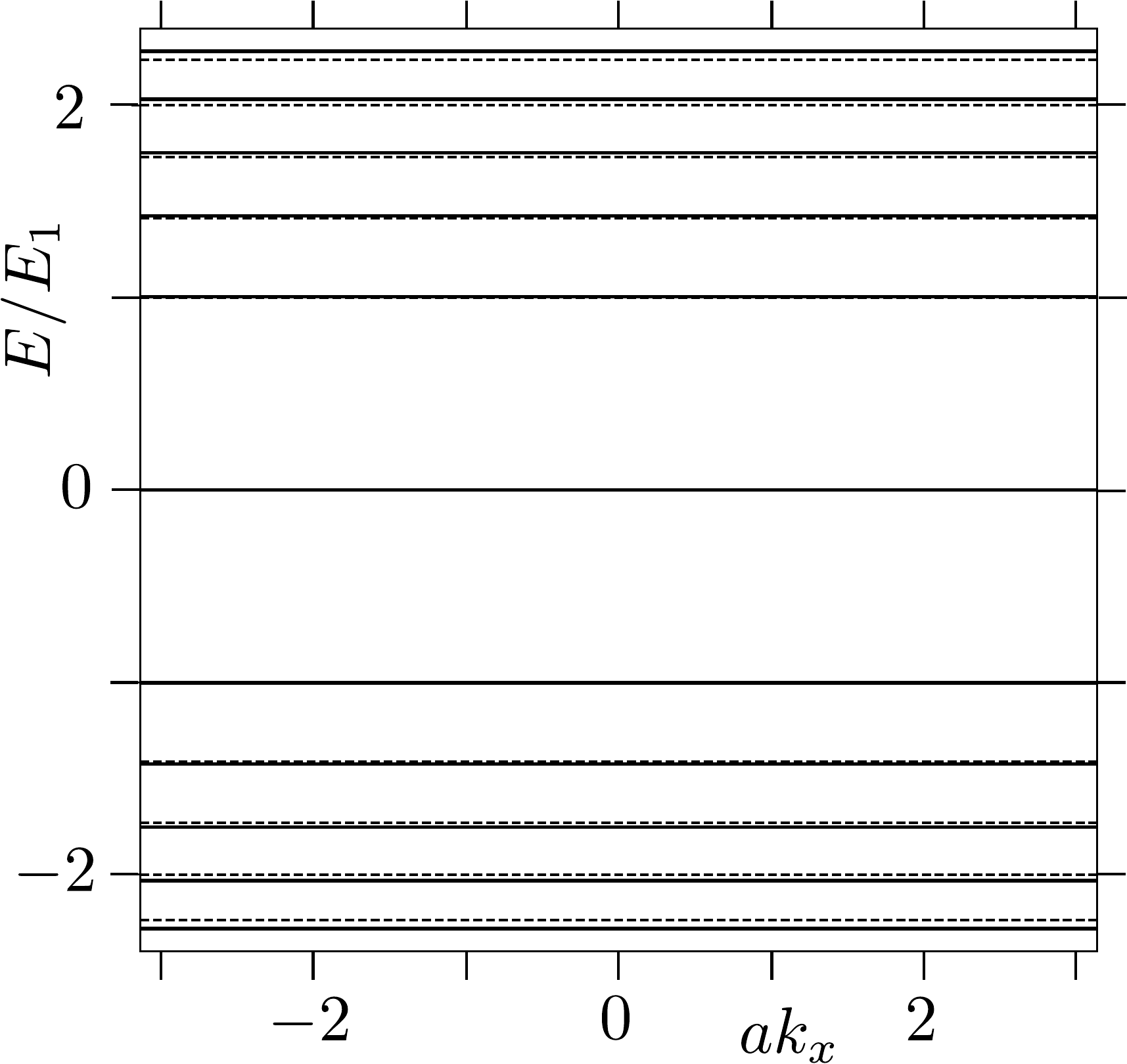}}
\caption{Solid lines: Dispersionless Landau levels in a uniform magnetic field $B_0=(1/201)(h/ea^2)$, calculated from the generalized eigenvalue equation \eqref{calHPhiPhidag} in the gauge $\bm{A}=-B_0y\hat{x}$. (Energies are plotted in units of $E_1=\sqrt{2\hbar ev^2 B_0}$.) The dashed lines indicate the continuum limit \eqref{Endef}. At each $k_x$-value there are two independent eigenstates in the zeroth Landau level, one with spin up and one with spin down. The other Landau levels each have only a single eigenstate at a given $k_x$, without any spin polarization.}
\label{fig_LL}
\end{figure}

In accord with these general considerations we have verified by explicit calculation (see Fig.\ \ref{fig_LL}) that the generalized eigenvalue problem \eqref{calHPhiPhidag} has an ${\cal N}$-fold degenerate zero-mode $E_0=0$ in both the $C=+1$ and $C=-1$ manifold.

To recover the chirality-resolved zeroth Landau we propose a method to spatially separate the opposite chirality manifolds: We double the system by adjoining a $+B$ and $-B$ region. Since then ${\cal I}=0$ by construction, the zeroth Landau level in each of the two regions could be chirality polarized without violating the no-go theorem. 

\begin{figure}[tb]
\centerline{\includegraphics[width=1\linewidth]{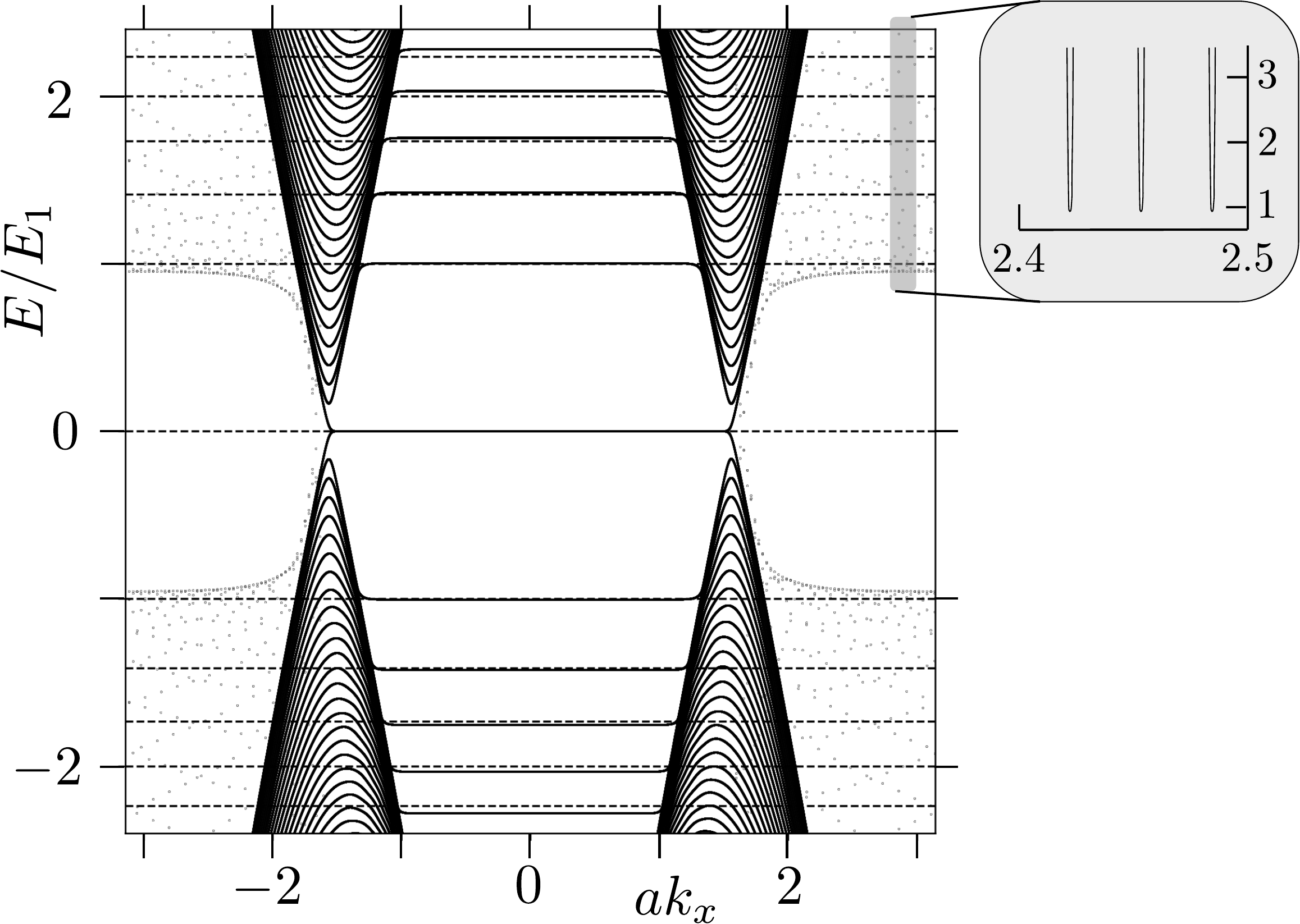}}
\caption{Same calculation as in Fig.\ \ref{fig_LL}, but for the non-uniform magnetic field profile with separate regions of $\pm B_0=\pm(1/202)(h/ea^2)$. The full profile of length $4L_0+a=405\,a$ is repeated periodically along the $y$-axis and is translationally invariant along the $x$-axis. The scattered data points near the Brillouin zone boundaries (with a nearly vertical dispersion, see expanded inset) are a lattice artefact.
}
\label{fig_zerothLL}
\end{figure}

\begin{figure}[tb]
\centerline{\includegraphics[width=0.8\linewidth]{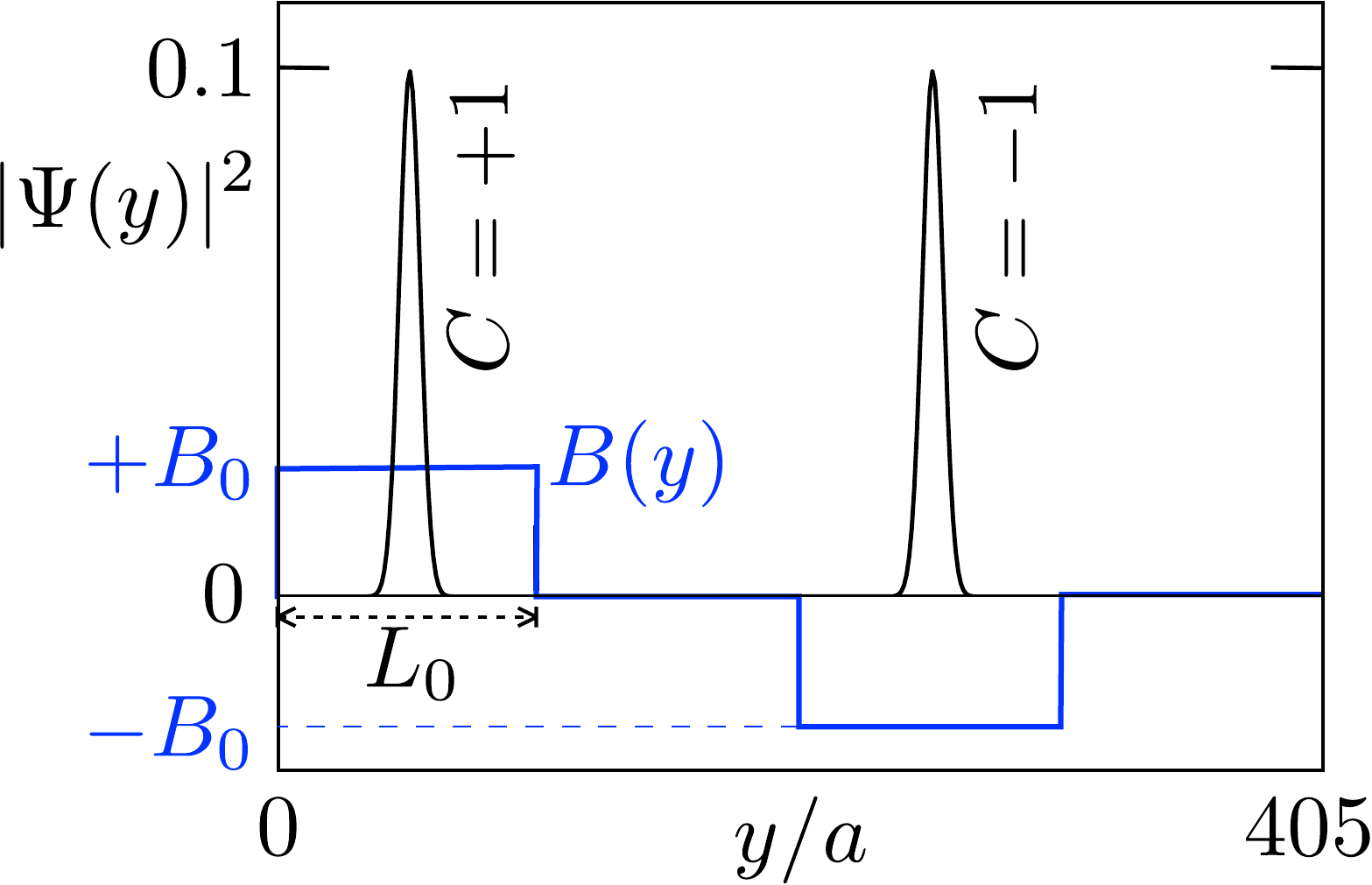}}
\caption{Wave function intensity profile in the zeroth Landau level for the band structure of Fig.\ \ref{fig_zerothLL}, evaluated at $k_x=0$. As indicated, the eigenstates with chirality $C=\pm 1$ (eigenvalue of $\sigma_z$) are spatially separated in the regions with magnetic field $\pm B_0$.
}
\label{fig_profile}
\end{figure}

Our numerics, see Figs.\ \ref{fig_zerothLL} and \ref{fig_profile}, shows that this is indeed what happens: the states in the zeroth Landau level with $C=\pm 1$ are fully contained in the $\pm B$ region.

In the next section we will check to what extent this spatial separation of the chiralities is sufficient to protect the flat band. 

\section{Robustness of the flat band}
\label{sec_robustness}

We introduce chirality-preserving disorder by randomly varying the perpendicular magnetic field component $B(x,y)$. The random field is drawn independently on each lattice site, uniformly in the interval $(0,2B_0)$ in the positive field region and in the interval $(-2B_0,0)$ in the negative field region.

\begin{figure}[tb]
\centerline{\includegraphics[width=0.75\linewidth]{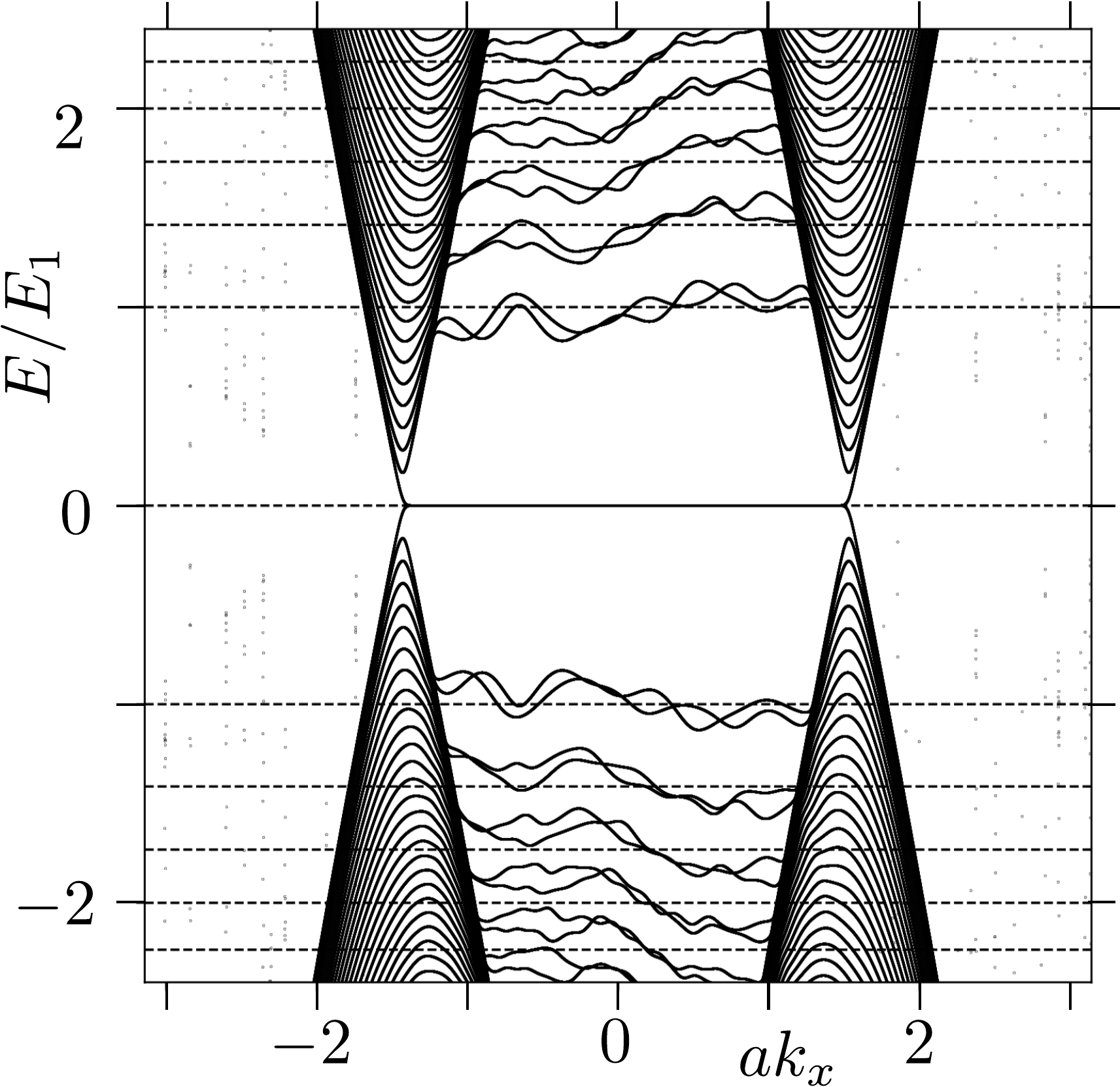}}
\caption{Same calculation as in Fig.\ \ref{fig_zerothLL}, but now for a magnetic field that varies randomly in the $y$-direction. The zeroth Landau level is protected from broadening because the states of opposite chirality are spatially separated.
}
\label{fig_disorder}
\end{figure}

\begin{figure}[tb]
\centerline{\includegraphics[width=0.75\linewidth]{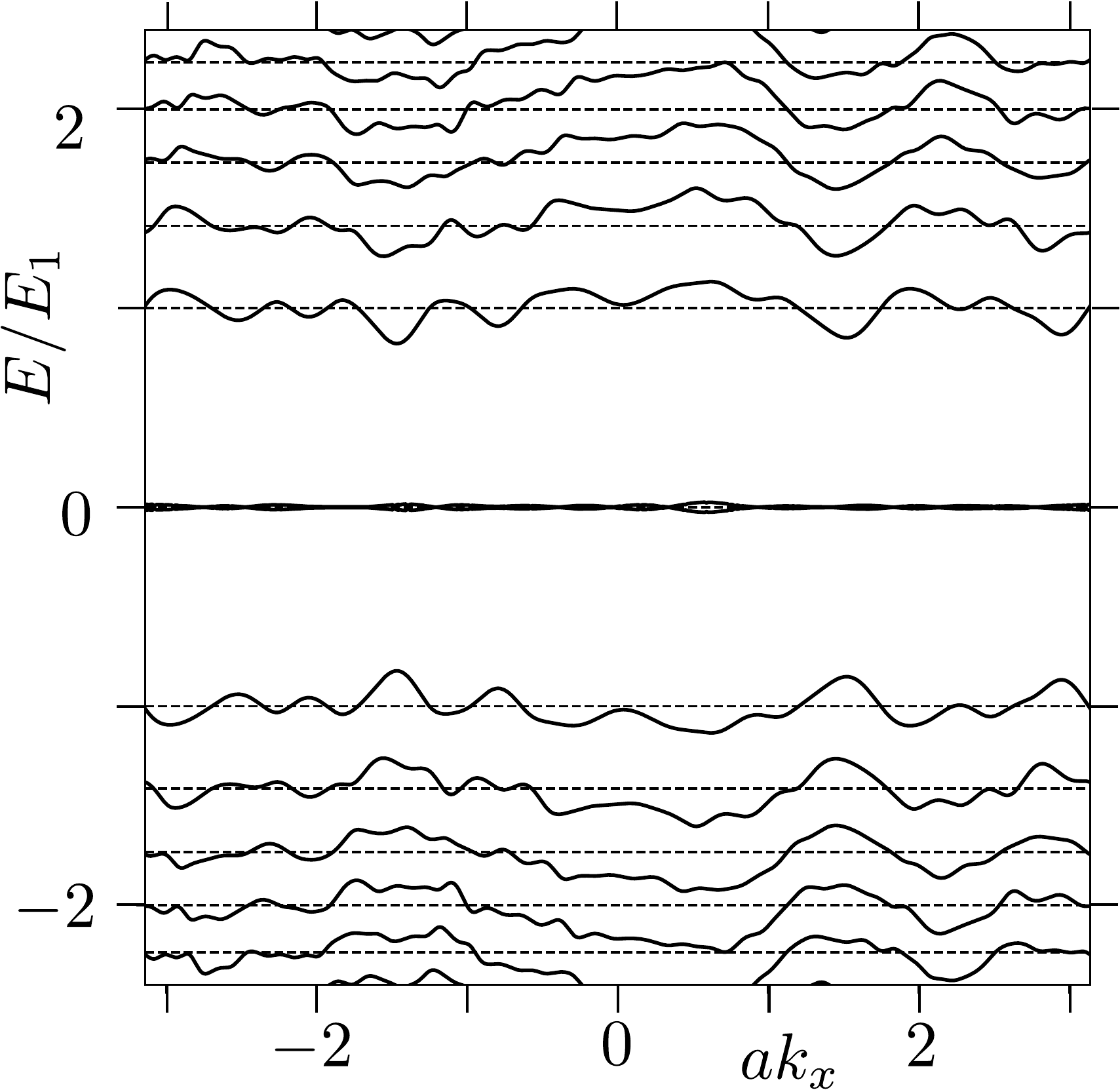}}
\caption{Same calculation as in Fig.\ \ref{fig_LL}, but now for a magnetic field that varies randomly in the $y$-direction. The zeroth Landau level contains states of opposite chirality which are not spatially separated, so they split in the presence of disorder.
}
\label{fig_disorder2}
\end{figure}

For the sake of illustration, it is helpful to first keep the translational invariance in the $x$-direction, so that $B(y)$ fluctuates only as function of $y$. We can then still plot a band structure as a function of $k_x$, see Fig.\ \ref{fig_disorder}. All flat bands are destroyed by the disorder, except for the zeroth Landau level, which remains completely dispersionless. The spatial separation of the states of opposite chirality is crucial for this topological protection: In Fig.\ \ref{fig_disorder2} we show that without it the zeroth Landau does broaden in the presence of disorder.

\begin{figure}[tb]
\centerline{\includegraphics[width=1\linewidth]{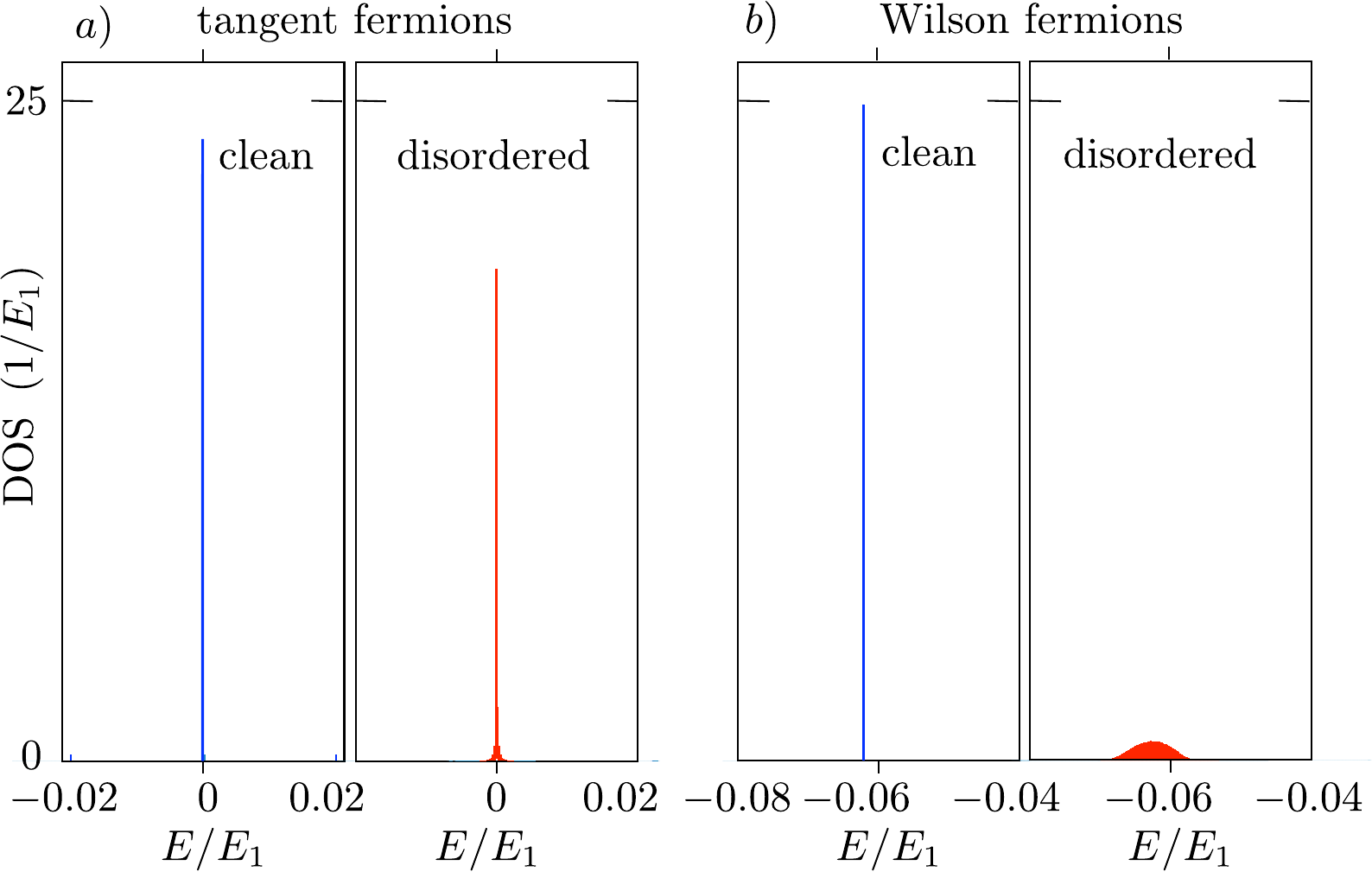}}
\caption{Density of states per unit cell for the tangent dispersion (a) and for the sine+cosine dispersion (b), with and without disorder in the magnetic field. (The disordered data is averaged over 1000 realizations.) The energy resolution is $2\cdot 10^{-4}\,E_1$, so that a peak of height $25/E_1$ corresponds to a degeneracy of 1 state per 200 unit cells. Both panels refer to the same magnetic field $B_0=(1/202)(h/ea^2)$, and the same disorder strength $B\in(0,2B_0)$. The geometry of panel a) is the $\pm B_0$ field profile of Fig.\ \ref{fig_profile} ($L_0=101\,a)$, while panel b) is for a single square of dimensions $202\,a\times 202\,a$. In both cases we impose periodic boundary conditions in the $x$-- and $y$--directions. The parameter $\Delta$ in the Wilson Hamiltonian \eqref{HWdef} is set at $\hbar v/a$.
}
\label{fig_DOS}
\end{figure}

We next consider a disordered field $B(x,y)$ that fluctuates in both $x$-- and $y$--directions. The wave number $k_x$ is then no longer a good quantum number, instead of a band structure we plot the density of states near $E=0$, to assess whether the zeroth Landau level is broadened. As shown in Fig.\ \ref{fig_DOS}a the density of states peak persists with only a slight broadening in the disordered system.

Earlier studies of the Landau level spectrum of lattice fermions use Wilson's sine+cosine dispersion \cite{Bru17,Bal18}, which breaks the chiral symmetry. In the Wilson Hamiltonian \eqref{HWdef} the zeroth Landau level is displaced from $E=0$ by the $\Delta$-dependent offset
\begin{equation}
\delta E=\tfrac{1}{2} eBa^2 \Delta/\hbar.\label{deltaEdef}
\end{equation}
The Wilson mass $\Delta$ is of order $\hbar v/a$ to effectively gap out the low energy excitations at $k=\pi/a$, hence $\delta E\simeq eBav$.

In Fig.\ \ref{fig_DOS}b we show results for the density of states, computed from the Wilson Hamiltonian for the same magnetic field value as in Fig.\ \ref{fig_DOS}a. Without disorder the only difference with the tangent dispersion is the shift \eqref{deltaEdef} of the zeroth Landau level, but with disorder the difference is quite dramatic.

\section{Conclusion}
\label{sec_conclude}

In summary, we have shown how the quantum Hall effect in a 3D topological insulator can be simulated on a 2D lattice. In a sense, the top and bottom surfaces in the slab geometry of Fig.\ \ref{fig_layout} are unfolded onto a plane. The inward and outward pointing magnetic field then corresponds to adjoining $+B$ and $-B$ regions, each with a zeroth Landau level of opposite chirality.

From a methodological point of view our work provides a gauge invariant way to discretize the Dirac equation on a lattice without breaking chiral symmetry. We note that earlier attempts to achieve this were not succesful \cite{Pot17,Don22a}. The defining equation \eqref{calHPhiPhidag} of tangent fermions has the form of a generalized eigenvalue problem, ${\cal H}\Psi=E{\cal P}\Psi$, with local Hermitian operators ${\cal H}, {\cal P}$ on both sides of the equations --- allowing for an efficient solution. 

The alternative method of Wilson fermions works with a conventional eigenvalue problem, $H_{\rm Wilson}\Psi=E\Psi$, that is local and gauge invariant, so it is certainly efficient. However, it breaks chiral symmetry, and it therefore lacks the topological protection of the zeroth Landau level. 

A previous study \cite{Don22b} has established the topological protection of the Dirac cone of tangent fermions in zero magnetic field. The present study completes this line of investigation by  showing how the topological protection can be extended to the zeroth Landau level in a magnetic field.

Our computer codes and numerical data are available at a repository, \doi{10.5281/zenodo.7495175}.

\section*{Acknowledgements}

We have benefited from discussions with M. Pacholski.\\
C.B. received funding from the European Research Council (Advanced Grant 832256).\\
J.T. received funding from the National Science Centre, Poland, within the QuantERA II Programme that has received funding from the European Union's Horizon 2020 research and innovation programme under Grant Agreement Number 101017733, Project Registration Number 2021/03/Y/ST3/00191, acronym {\sc tobits}.

\end{document}